\documentclass[aps,prl,twocolumn
,groupedaddress,showkeys]{revtex4-2}

\usepackage{amsmath}
\usepackage{amssymb}
\usepackage{xcolor}
\usepackage{tikz} 
\usepackage{pgfplots}
\usetikzlibrary {arrows.meta}
\usetikzlibrary{shapes.geometric}
\usetikzlibrary{bending}
\usetikzlibrary{pgfplots.fillbetween}
\usepackage{hyperref}
\usepackage[justification=justified]{subcaption}
\usepackage{graphicx}
\usepackage[font=small,labelfont=bf,
   justification=justified]{caption}

\newcommand{\be}{\mathbf{e}}

\newcommand{\bw}{\mathbf{w}}
\newcommand{\bx}{\mathbf{x}}
\newcommand{\by}{\mathbf{y}}
\newcommand{\bz}{\mathbf{z}}

\begin{document}

\title{Locality approach to the bootstrap percolation paradox}

\author{Ivailo Hartarsky}
\email{ivailo.hartarsky@tuwien.ac.at}
\affiliation{Technische Universit\"at Wien, Institut für Stochastik und Wirtschaftsmathematik, Wiedner Hauptstra\ss e 8-10, A-1040, Vienna, Austria}
\author{Augusto Teixeira}
\email{augusto@impa.br}
\affiliation{IMPA, Estrada Dona Castorina, 110 - Rio de Janeiro, Brazil}

\date{\today}

\begin{abstract}
We revisit the Bootstrap Percolation model, leveraging recent mathematical advances linking it with its local counterpart. This new perspective resolves, for the first time, historic discrepancies between Monte Carlo simulations and theoretical results: previously, those predictions disagreed even in the first-order asymptotics of the model. In contrast, our framework achieves excellent agreement between numerics and theory, which now match up to the third-order expansion, as the infection probability approaches zero. Our algorithm allows us to generate novel predictions for the model.
\end{abstract}

\maketitle

Monte Carlo simulations provide an accessible tool for probing physical systems and have been used extensively. Yet, the reliability of results based on such approaches is not guaranteed. Along with explosive percolation \cite{Achlioptas09,Riordan11,Li23}, bootstrap percolation (BP) is among the prime examples of how Monte Carlo simulations can yield misleading results. Indeed, for BP, going back to \cite{Pollak75,Chalupa79,Kogut81}, this phenomenon is so common that is has been labelled the \emph{BP paradox} \cite{DeGregorio04} (also see \cite{VanEnter90,Adler03,Gravner08,DeGregorio09}). BP is a paradigmatic statistical mechanics model of metastability with deep connections to the low temperature stochastic Ising model \cite{Dehghanpour97a,Morris11,Cerf13}, kinetically constrained models of supercooled liquids \cite{Ritort03,Cancrini08,Garrahan11,Arceri21} and the $k$-core model \cite{Guggiola15}, as well as applications to areas as diverse as complexity theory \cite{Goles13}, influence in social networks \cite{Banerjee20}, sandpile formation \cite{Manna98} and many others. We direct the reader to the surveys \cite{Adler91,DeGregorio09,Morris17,Morris17a} for more background on BP.

The history of the BP paradox can be summarised as follows: each prediction of asymptotic behaviour based on numerics was subsequently proved erroneous by rigorous results. The list is impressive: triviality of the phase transition \cite{Kogut81,VanEnter87,Adler88,Schonmann92}, critical scaling \cite{Aizenman88,Cerf02,Adler03,Kurtsiefer03}, sharp threshold \cite{Lenormand84,Nakanishi86,Adler89,Adler03,Holroyd03,Holroyd06}, second order critical exponent \cite{Adler03,DeGregorio05,DeGregorio06,Gravner08,DeGregorio09,Gravner09,Gravner12} and second order logarithmic corrections \cite{Teomy14,Hartarsky19} have all gone through this pattern over the years. Two decades ago \cite{DeGregorio04,DeGregorio05,DeGregorio06}, an important attempt at reconciling theory and numerics was made. Yet, this led to more discrepancies \cite{DeGregorio09,Gravner09} and ultimately only widened the gap between communities: ``The striking conclusion is that not even the rigorous correction term can be captured reliably by the numerical exact solution yet'' \cite{DeGregorio09}. Following our recent progress on the rigorous theory side \cite{Hartarsky24locality} and, presently, the numerics side, we are now able to settle this recurrent controversy, explain its origin and, finally, show that it is possible to make accurate predictions for these models based on numerical results.

\section{Models}
We will discuss two closely related BP models on the square lattice $\mathbb Z^2$. Start with a set $A_0\subset\mathbb Z^2$ of initially infected sites (also known as + spins, vacancies, etc.\ in other contexts) with density $p\in(0,1)$, selected independently at random. At each discrete time step we define $A_t$ by adding to $A_{t-1}$ all sites $\bx\in\mathbb Z^2$ fulfilling the following condition.
\begin{itemize}
    \item {\bf MBP} - Modified two-neighbour BP:
    there is at least one infection in $A_{t-1}$ among both the horizontal neighbours $\{\bx-\be_1,\bx+\be_1\}$ and the vertical neighbours $\{\bx-\be_2,\bx+\be_2\}$ of $\bx$ (e.g.
    \begin{tikzpicture}[scale=0.1]
        \draw[color=white!30!black] (0, 0) rectangle (1, 1);
        \draw[color=white!30!black, fill=black] (0, 1) rectangle (1, 2);
        \draw[color=white!30!black, fill=black] (1, 0) rectangle (2, 1);
        \draw[->] (2.6, 1) -- (4.4, 1);
        \draw[color=white!30!black, fill=red!70!black] (5, 0) rectangle (6, 1);
        \draw[color=white!30!black, fill=black] (5, 1) rectangle (6, 2);
        \draw[color=white!30!black, fill=black] (6, 0) rectangle (7, 1);
    \end{tikzpicture}
    ).
    \item {\bf FBP} - Frob\"ose BP:
    there is a four-cycle $\bx\by\bz\bw$ such that $\by,\bz,\bw$ are all infected in $A_{t-1}$
    (e.g.
    \begin{tikzpicture}[scale=0.1]
        \draw[color=white!30!black] (0, 0) rectangle (1, 1);
        \draw[color=white!30!black, fill=black] (0, 1) rectangle (1, 2);
        \draw[color=white!30!black, fill=black] (1, 0) rectangle (2, 1);
        \draw[color=white!30!black, fill=black] (1, 1) rectangle (2, 2);
        \draw[->] (2.6, 1) -- (4.4, 1);
        \draw[color=white!30!black, fill=red!70!black] (5, 0) rectangle (6, 1);
        \draw[color=white!30!black, fill=black] (5, 1) rectangle (6, 2);
        \draw[color=white!30!black, fill=black] (6, 0) rectangle (7, 1);
        \draw[color=white!30!black, fill=black] (6, 1) rectangle (7, 2);
    \end{tikzpicture}
    ).
\end{itemize}
We remark that the classical two-neighbour constraint can be treated like FBP \cite{Hartarsky24locality}, but is omitted for concision.

\begin{figure*}
    \centering
    \begin{subcaptionblock}[t]{0.5\textwidth}
        \centering
        \includegraphics[width=8.6cm]{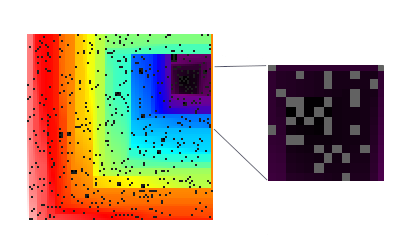}
        \caption{\label{fig:droplet:FBP}FBP: infection probability $0.06$, side length $93$.}
    \end{subcaptionblock}%
    ~ 
    \begin{subcaptionblock}[t]{0.5\textwidth}
        \centering
        \includegraphics[width=8.6cm
        ]{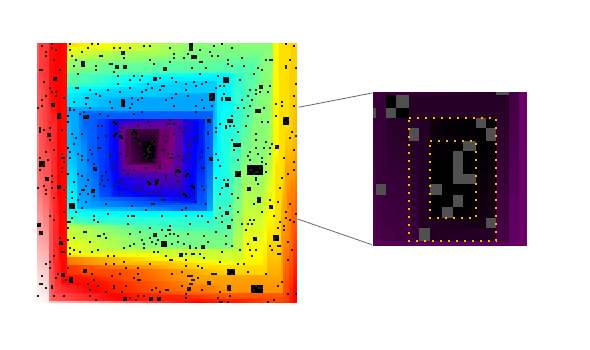}
        \caption{\label{fig:droplet:MBP}MBP: infection probability $0.047$, side length $130$.}
    \end{subcaptionblock}
    \caption{\raggedright Critical droplets conditioned to be internally filled. Colors represent infection times, with grey being the initial infections.
    The little windows contain a zoomed-in picture of the early stages of nucleation, with initial infections in lighter grey.
    Close inspection reveals that local FBP internally fills with the same initial condition, while local MBP does not. The dotted rectangles on the right show two moments where local MBP becomes confined.}
    \label{fig:droplet}
\end{figure*}

Since, in infinite volume, BP exhibits a trivial phase transition with $p_{\mathrm c}=0$ \cite{VanEnter87}, we focus on the low temperature regime $p\to 0$. As it is common in metastability, key quantities in BP can be expressed in terms of the frequency (density) $\rho$ of occurrence of certain critical droplets enabling nucleation \cite{Aizenman88}. In our models, a critical droplet is simply an axis-parallel square of side length $\Lambda=C\log(1/p)/p$ (with $C$ an appropriately large constant, for us $C=2$ suffices) which is \emph{internally filled}: the initial infections of $A_0$ inside the square are sufficient to infect it completely, see Fig.~\ref{fig:droplet}.

\section{Background}
It is useful to recall previous approaches for estimating the critical droplet density $\rho$. The most natural one consists in running a Monte Carlo simulation of a square of side $\Lambda$ with initial infection density $p$ and record the empirical proportion of samples infecting the entire square. The main drawback of this straightforward approach, which has been extensively used \cite{Kogut81,Lenormand84,Nakanishi86,Adler88,Adler89,Adler03,Kurtsiefer03}, is that its running time is governed by $\rho^{-1}$ rather than $\Lambda$. In metastability settings like ours, this leads to an exponential complexity, since $\rho^{-1}\sim\exp(C'/p)$ \cite{Aizenman88}, as we will see. This practically limits Monte Carlo simulations to moderate densities $p>0.05$, see Fig.~\ref{fig:plots}. Despite the significant computational power needed to obtain such results, extrapolating them to predict thermodynamic results is risky.

An approach to go beyond moderate densities was proposed in \cite{DeGregorio04,DeGregorio05,DeGregorio06}, based on a local version of BP. This simplified model \cite{Gravner09} prohibiting polynucleation is defined as follows. Consider a growing sequence of nested rectangles $R_n$ as follows. The seed $R_0$ is a single initially infected site. At each step, $R_{n+1}$ is obtained from $R_n$ by increasing the height or width by 1, but we require that the additional line (row or column) contains at least one initial infection. For local MBP we also allow $R_{n+1}$ to be obtained from $R_n$ by adding both a row and a column to $R_{n}$, provided the corner at their intersection is initially infected. The local density $\rho_\ell$ is the probability that, starting at $R_0=\{\mathbf{0}\}$, there exists such a sequence reaching a square of side $\Lambda$. The difference with the original BP is that we do not allow multiple non-trivial rectangles to first grow and then merge.

Local BP is much more tractable, because the probability of reaching a rectangle of given size can be computed recursively \cite{DeGregorio06,Hartarsky24locality}. In order to do this, we need to record some additional information on which sides of the current rectangle we have already attempted and failed to extend. Yet, there are only a finite number of such `frame states' (seven for MBP and eight for FBP \cite{Hartarsky24locality}), so these probabilities can be computed recursively. This yields a dynamic programming algorithm for computing $\rho_\ell$ with time and memory complexities of $\Lambda^2$ and $\Lambda$ respectively. As opposed to the exponential cost of Monte Carlo simulations, this local approach is polynomial in $1/p$. This allowed probing densities down to $p\approx0.0035$ \cite{DeGregorio05} for local MBP with minimal computational resources. Moreover, the output of this algorithm is deterministic and only subjected to numerical precision errors, which can be controlled to arbitrary precision.

Strikingly, local BP computations still led to incorrect predictions of the asymptotic scaling of $\rho_\ell$ for MBP as $p\to 0$ \cite{Gravner09}, even when informed of the rigorous result $\rho\sim\rho_\ell\sim\exp(-\pi^2/(3p))$ \cite{Holroyd03}. The disappointing results of this very promising method essentially extinguished the hope of numerical and rigorous results reaching agreement \cite{DeGregorio09}. Consequently, the numerical side of the area has seen no activity in the last decade, since the last incorrect predictions of \cite{Teomy14}, based on a version of local BP, later refuted in \cite{Hartarsky19}. During the same period, the rigorous theory has been flourishing with a number of very precise results for specific models \cite{Balogh12,Duminil-Copin13,Bollobas17,Duminil-Copin24,Hartarsky23FA,Hartarsky24locality} and the development of a unified universality theory for BP and the closely related kinetically constrained models \cite{Martinelli19a,Mareche20Duarte,Balister22,Hartarsky22univlower,Bollobas23,Balister24,Hartarsky24univupper,BalisterNaNb}.

\begin{figure*}
    \centering
\begin{subcaptionblock}[t]{0.5\textwidth}
    \centering
    \begin{tikzpicture}[x=0.63cm,y=3cm]
\draw[->,color=black] (1,0.25) -- (12.5,0.25) node[above]{$\log\frac1p$};
\foreach \x in {1,2,3,4,5,6,7,8,9,10,11,12}
\draw[shift={(\x,0.25)}] (0pt,2pt) -- (0pt,-2pt) node[below] {\footnotesize $\x$};
\draw[->,color=black] (1,0.25) -- (1,1.75) node[right]{$-p\log\rho_\ell$, {\color{blue}$-p\log(p^2\rho)$}};
\foreach \y in {0.5,1,1.5,2,2.5,3}
\draw[shift={(1,0.5*\y)}] (2pt,0pt) -- (-2pt,0pt) node[left] {\footnotesize $\y$};
\fill (1.38629436111989076, 0.4557867386130542) circle (1pt);
\fill (2.0794415416798357, 0.5928339472416244) circle (1pt);
\fill (2.7725887222397811, 0.7745581895375311) circle (1pt);
\fill (3.4657359027997265, 0.9546038641712591) circle (1pt);
\fill (4.1588830833596715, 1.1129858612276886) circle (1pt);
\fill (4.8520302639196169, 1.2430073481077526) circle (1pt);
\fill (5.5451774444795623, 1.345804448976588) circle (1pt);
\fill (6.2383246250395077, 1.424783934531369) circle (1pt);
\fill (6.9314718055994531, 1.4842946522264553) circle (1pt);
\fill (7.6246189861593985, 1.5284964038060636) circle (1pt);
\fill (8.317766166719343, 1.5609736388102249) circle (1pt);
\fill (9.0109133472792884, 1.5846389054119723) circle (1pt);
\fill (9.7040605278392338, 1.6017726607118656) circle (1pt);
\fill (10.397207708399179, 1.6141156659249392) circle (1pt);
\fill (11.090354888959125, 1.6229727530039586) circle (1pt);
\fill (11.78350206951907, 1.6293089682631352) circle (1pt);
\draw[color=blue] (1.3862943611198906, .5 * 1.1088918504004779
) circle (1pt);
\draw[color=blue] (1.5249237972318797, .5 * 1.0791886883428758
) circle (1pt);
\draw[color=blue] (1.6635532333438687, .5 * 1.0681367810119331
) circle (1pt);
\draw[color=blue] (1.8021826694558578, .5 * 1.0866606233031444
) circle (1pt);
\draw[color=blue] (1.940812105567847, .5 * 1.094153392261219
) circle (1pt);
\draw[color=blue] (2.0794415416798357, .5 * 1.1603930796913482
) circle (1pt);
\draw[color=blue] (2.2180709777918253, .5 * 1.2286592484438938
) circle (1pt);
\draw[color=blue] (2.3567004139038144, .5 * 1.2764276865707216
) circle (1pt);
\draw[color=blue] (2.495329850015803, .5 * 1.3524372314450723
) circle (1pt);
\draw[color=blue] (2.633959286127792, .5 * 1.4258671562505334) circle (1pt);
\draw[color=blue] (2.772588722239781, .5 * 1.4990440977520134) circle (1pt);

\draw (1,1.644934066848226436472415166646) node[left]{$\pi^2/3$}--(12,1.644934066848226436472415166646);
\draw[domain=3:12,smooth] plot (\x,{1.644934066848226436472415166646-5.8049063042788620281153757483082*exp(-\x/2)});
\draw (4.3,0.75) node[right]{$\frac{\pi^2}3-2\pi\sqrt{2+\sqrt2}\sqrt p$};
\draw[domain=1.3862943611198906:12,smooth,dashed,red] plot (\x,{1.644934066848226436472415166646-5.8049063042788620281153757483082*exp(-\x/2)+5.0765*exp(-4*\x/5)});
\draw (5,1.1) node[right]{\color{red}$\frac{\pi^2}3-2\pi\sqrt{2+\sqrt2}\sqrt p+10.153p^{4/5}$};
\end{tikzpicture}
    \caption{\raggedright FBP: The solid line and curve are the rigorous first and second order asymptotics. The estimated third order is dashed.}
    \label{fig:FBP}
\end{subcaptionblock}%
~
\begin{subcaptionblock}[t]{0.5\textwidth}
    \centering
    \begin{tikzpicture}[x=0.63cm,y=0.34cm]
\draw[->] (1,2) -- (12.5,2) node[above left]{$\log\frac1p$};
\foreach \x in {1,2,3,4,5,6,7,8,9,10,11,12}
\draw[shift={(\x,2)}] (0pt,2pt) -- (0pt,-2pt) node[below] {\footnotesize $\x$};
\draw[->] (1,2) -- (1,15) node[right]{$\frac{\pi^2}{3\sqrt p}+\sqrt p\log\rho_\ell$, {\color{blue}$\frac{\pi^2}{3\sqrt p}+\sqrt p\log(p^2\rho)$}};
\foreach \y in {4,8,12,16,20,24,28}
\draw[shift={(1,0.5*\y)}] (2pt,0pt) -- (-2pt,0pt) node[left] {\footnotesize $\y$};
\fill (1.3862943611198906, 2.5336173824903687) circle (1pt);
\fill (2.0794415416798357, 3.4542750157635616) circle (1pt);
\fill (2.7725887222397811, 4.375715428902341) circle (1pt);
\fill (3.4657359027997265, 5.269321110664854) circle (1pt);
\fill (4.1588830833596715, 6.127509797976364) circle (1pt);
\fill (4.8520302639196169, 6.950806312754377) circle (1pt);
\fill (5.5451774444795623, 7.742799297258774) circle (1pt);
\fill (6.2383246250395077, 8.507654470088749) circle (1pt);
\fill (6.9314718055994531, 9.24951321142202) circle (1pt);
\fill (7.6246189861593985, 9.972215649531723) circle (1pt);
\fill (8.317766166719343, 10.679135394361651) circle (1pt);
\fill (9.0109133472792884, 11.373169080311992) circle (1pt);
\fill (9.7040605278392338, 12.056750506103697) circle (1pt);
\fill (10.397207708399179, 12.731894171509936) circle (1pt);
\fill (11.090354888959125, 13.400250258565961) circle (1pt);
\fill (11.78350206951907, 14.063159033781114) circle (1pt);
\draw[color=blue] (1.3862943611198906, 0.5 * 4.7504838951308681) circle (1pt);
\draw[color=blue] (1.5249237972318797, 0.5 * 5.1917104101223526) circle (1pt);
\draw[color=blue] (1.6635532333438687, 0.5 * 5.5862617804263355) circle (1pt);
\draw[color=blue] (1.8021826694558578, 0.5 * 6.0734435304747265) circle (1pt);
\draw[color=blue] (1.9408121055678467, 0.5 * 6.6400832304974813) circle (1pt);
\draw[color=blue] (2.0794415416798357, 0.5 * 7.1398747365457789) circle (1pt);
\draw[color=blue] (2.2180709777918248, 0.5 * 7.6140431639373949) circle (1pt);
\draw[color=blue] (2.3567004139038139, 0.5 * 8.091141532501668) circle (1pt);
\draw[color=blue] (2.495329850015803, 0.5 * 8.5390975964028915) circle (1pt);
\draw[color=blue] (2.6339592861277921, 0.5 * 8.9436345088255482) circle (1pt);
\draw[color=blue] (2.7725887222397811, 0.5 * 9.3808186041235047) circle (1pt);
\draw[color=blue] (2.9112181583517702, 0.5 * 9.7268053067624134) circle (1pt);
\draw[domain=1:12,smooth] plot (\x,{3.2695+0.92388*\x});
\draw (9,12) node[left]{$\sqrt{2+\sqrt2}\log\frac1p+6.539$};
\draw[domain=1:12,smooth,dashed,red] plot (\x,{0.92388*\x+3.2695-3.72*exp(-0.31*\x)});
\draw (2.8,4) node[right]{\color{red}$\sqrt{2+\sqrt2}\log\frac1p+6.539-7.44p^{0.31}$};
\draw[dotted] (1,2) rectangle (5.7,8);
\end{tikzpicture}
    \caption{\raggedright MBP: The solid line depicts the second and third order terms. The dashed curve incorporates the estimated fourth order correction. The dotted box indicates the previously explored range \cite{DeGregorio06}.
    \label{fig:MBP}}
\end{subcaptionblock}
\caption{\raggedright \label{fig:plots}Exact numerical values of the local critical droplet density $\rho_\ell$ for $p = 2^{-k}$ with $k\in\{2,\dots,17\}$.
Empty circles represent Monte Carlo simulation data for $p^2\rho$ instead of $\rho_\ell$. 
The confidence intervals are smaller than the circles used in the figure.}
\end{figure*}

\section{Rigorous locality approach}
The present letter is largely inspired and complemented by our recent mathematical work \cite{Hartarsky24locality}.
It is our goal to elucidate the new perspective it offers along with its implications on the rigorous and numerical side of the BP problem. Let us start by stating one of our main results \cite[Theorem~1.1, Lemma~3.5]{Hartarsky24locality}, the locality theorem. For FBP, the non-locality $\log\rho-\log \rho_\ell$ is at most polylogarithmic in $1/p$ \footnote{In \cite{Hartarsky24locality}, the polylogarithmic bound on non-locality features the large exponent 20. This is only for the sake of simplifying the proof. We do not expect this exponent to be large. Moreover, in view of Fig.~\ref{fig:FBP}, it is plausible that non-locality may be $2\log(1/p)+o(\log(1/p))$.}. This establishes in a very strong sense that $\rho\sim\rho_\ell$. We also provide compelling numerical support for this fact in Figs.~\ref{fig:droplet:FBP} and \ref{fig:FBP}. While previously local FBP was introduced as a simplification of FBP, it is now clear that these two models are essentially equivalent. The locality theorem is important to both theory and numerics, because the local model is much more accessible. Owing to locality, we could prove \cite[Theorem~1.2]{Hartarsky24locality} that, for FBP,
\begin{equation}
\label{eq:FBP}
\log\rho\sim\log\rho_\ell\sim\frac{-\pi^2}{3p}+\frac{2\pi\sqrt{2+\sqrt 2}}{\sqrt p}.\end{equation}
At present, we do not seek to discuss the proof of \eqref{eq:FBP} (see \cite{Hartarsky24locality}), but rather the implications of the locality theorem to numerics, keeping \eqref{eq:FBP} as a verification reference.

Since the discrepancy between the local and non-local infection times is much smaller than the precision currently feasible for $\log\rho$, the numerical approach to local BP in fact yields precise results also for the non-local model. Thus, it is all the more important to understand the reasons for its previous failure to predict the critical exponent $1/2$ of the second term in \eqref{eq:FBP}, even given the first order term. In order to do this, we have performed a contemporary implementation of the dynamical programming algorithm. While this is not a difficult task, some care is needed to obtain an efficient implementation. 

There are two key novel features in our simulations.
First, we take advantage of the parallel nature of the computation through the use of a GPU.
Also, we do not store the desired probabilities directly and take care or error propagation when operating over them. This is necessary, since the probabilities $\rho_\ell$ that we attain are as small as $10^{-2\cdot10^5}$ and some off-diagonal values can be as small as $p^{-\Lambda}\approx10^{-2700000}$.
So a trade-off is needed between numerical precision and efficiency.
See more details in the supplemental material.
We alert the reader that, even though such extraordinarily small numbers have little meaning in physical context, the results they will be used to obtain do have tangible implications in regimes of physical interest.

Running the algorithm discussed above and available at the repository \url{https://github.com/augustoteixeira/bootstrap} (see folder \url{taichi}), we are able to reach $p\approx8\cdot10^{-6}$. The resulting values of $p\log \rho_\ell$ are presented in Fig.~\ref{fig:plots} (see the supplemental material for the numerical values). For comparison, we have also implemented the straightforward Monte Carlo strategy (see supplemental material), displaying one sample internally filled critical droplet in Fig.~\ref{fig:droplet} and data for $p^2 \rho$ in Fig.~\ref{fig:plots}. A detailed explanation for the corrective factor $p^2$ is provided in the supplemental material, but it is necessary for roughly compensating the volume $\Lambda^2$ of the droplet and its contribution is only relevant for the large and moderate values of $p$ accessible to Monte Carlo simulations.

In the words of \cite{DeGregorio06}, ``If we are to interpret quantitative data reliably, there is
now a pressing need for advancement of the whole theoretical agenda.'' Indeed, efficient data analysis is a crucial and novel point of our work. Namely, we develop a reliable scheme for extracting extremely precise asymtptotics and determining their correct functional form. The details are provided in the supplemental material, since we expect such treatment to be of use well beyond the scope of BP. But let us present our results obtained by following these guiding principles: allow general functional forms rather than artificially imposing power laws; use a wide range of data to roughly estimate the correct functional form of the subsequent error term before estimating the main one; use only the latest data for precise estimates, given the functional form.

As an illustration, consider estimating the limit $\pi^2/3\approx3.28987$ from the data in Fig.~\ref{fig:FBP}. What one should \emph{not} do is to take the value of the last data point. In our case, this gives $3.2586$ for FBP (and even $3.2122$ for MBP). Instead, we use the last few points to infer the functional form of the leading and second order terms. Once this is done, performing a 3-parameter fit of the form $p\log\rho_\ell^{-1}\sim a+bp^{-c}$, using the last four data points, gives $a\approx3.290(3)$: four correct digits instead of one or two. Proceeding similarly, we are able to infer the functional form in \eqref{eq:FBP} and the constants appearing, as well as the third order term, which is not known rigorously. For the second order constant $2\pi\sqrt{2+\sqrt 2}$, we also obtain three correct digits. We estimate the third order exponent for FBP to be approximately $0.19(6)$ based only on the last two data points. Assuming it is $1/5$ and using only the last data point, we obtain the remarkable 1-parameter fit depicted in Fig.~\ref{fig:FBP}. The reason not to consider further corrective terms is that, strikingly, the error $|\log\rho_\ell^{-1}-\pi^2/(3p)+2\pi\sqrt{2+\sqrt2}/\sqrt p-10.153/\sqrt[5]p|$ is less than $2$ throughout the entire range $\log p^{-1}\in[1,12]$ and actually decays in the second half of this interval. In this interval, $\log\rho_\ell^{-1}$ increases from $2$ to $5\cdot10^5$. This suggests that this might be the final divergent term in the asymptotics of $\log\rho_\ell^{-1}$ for FBP. This is supported, for instance by results in simpler but somewhat similar settings \cite{ElveyPrice21}, though the value of this exponent did not match our heuristic expectations.

A similar analysis allows us to detect the asymptotics
\begin{equation}
\label{eq:MBP}
\log\rho_\ell\sim\frac{-\pi^2}{3p}+\frac{\sqrt{2+\sqrt 2}\log(1/p)}{\sqrt p}+\frac{6.54}{\sqrt p}\end{equation}
for local MBP, very clearly distinguishing the logarithmic correction (see Fig.~\ref{fig:MBP}). Indeed, it is essential to consider richer functional expressions to obtain accurate results for this model \cite{Hartarsky23mod-2n}. In this case, we do expect at least one subsequent term with exponent around $0.19$ to be neglected in \eqref{eq:MBP}. For reference, the first order term in \eqref{eq:MBP} is known rigorously \cite{Holroyd03}, the second one was recently conjectured \cite[Section~8.2]{Hartarsky24locality} based on theoretic considerations, while the third and subsequent order terms are new.

\section{What went wrong?}
While the above results are rather satisfactory, for the sake of other models, it is important to reflect on what makes this method work, as opposed to the original local BP approach in \cite{DeGregorio04,DeGregorio05,DeGregorio06}. Indeed, as far as the algorithm is concerned, disregarding implementation, the two are effectively identical. Yet, \cite{DeGregorio06} mistakenly predicted a second term of order $p^{-2/3}$, while we achieve perfect agreement even to third order, based on the same kind of data. 

So what went wrong twenty years ago? In fact, several things. Firstly, in the present approach, we do not artificially limit numerical computations to physically relevant scales. This may initially seem like a bad idea. However, we saw that even the last few data points, all corresponding to system sizes well beyond the size of the universe, allow us to reliably extract very precise information, valid down to very small system sizes and consistent with rigorous asymptotic results previously accused of being irrelevant at physically meaningful scales \cite{Adler03}. For this reason it is essential that we take full advantage of the efficiency of the algorithm and its implementation.
In other words, by considering large (and un-physical) scales, we effectively obtain more terms in the expansion of $\log{\rho}$ around infinity, which are crucial to understanding medium sized (and physically relevant) systems.

Secondly, in view of \eqref{eq:FBP} and \eqref{eq:MBP}, one should allow some richness in the functional form to be fitted. Indeed, at our values of $p$, a logarithmic factor can easily be mistaken for a small power, if we do not take into account several data points to infer the correct shape. This is why, in \cite{DeGregorio06}, a second order exponent for MBP of $2/3$ was estimated, instead of $1/2$ with a logarithmic correction. It is thus important to consider fairly general expressions in order to test what kind of terms are actually needed, rather than artificially imposing a power law.

Finally, let us mention one last important aspect of the problem, for which the theoretical understanding at the time of \cite{DeGregorio06} was not yet ripe. The locality theorem stated above implies that the non-locality $\log \rho-\log\rho_\ell$ is very small for FBP. However, for MBP, we expect this quantity to be much larger, of order $p^{-1/2+o(1)}$ \cite[Section~8.2]{Hartarsky24locality}, as also supported by Figs.~\ref{fig:droplet:MBP} and~\ref{fig:MBP}. Thus, going beyond the precision of \eqref{eq:MBP} for local MBP will not lead to more accurate results for non-local MBP. 
Hence, for models other than FBP and two-neighbour BP, for which locality is proved in \cite{Hartarsky24locality}, it is crucial to assess the validity and precision of the local approximation rather than applying it directly, as done in \cite{DeGregorio04,Teomy14}.

\section{Conclusion}
We proposed a new viewpoint on BP based on rigorous quantitative locality. Local BP enabled us to make progress both on the rigorous theory and on the numerics of the problem. For the former, we proved exact second order asymptotics for FBP and two-neighbour BP, thanks to the strong locality theorem for these models. Concerning numerical methods, we have revived the approach pioneered in \cite{DeGregorio04} for MBP, extended it to other models and implemented it efficiently. This allowed us to obtain a simple functional form in remarkable agreement with rigorous asymptotics, Monte Carlo simulations, in the regime where they are feasible, and exact numerical computations of the local density of critical droplets. We identified the caveats associated with the BP problem and developed a reliable methodology for avoiding them. This enabled a detailed understanding of the shortcomings of previous methods.

While the study of FBP and two-neighbour BP appears to be near its end, following this progress, this is far from being the case for other BP and related models, particularly those relevant for understanding glassy dynamics. It is therefore important that we have developed a detailed understanding of a few paradigmatic models, which can be used for bench-marking and calibrating future theoretical and numerical investigations. 

In conclusion, let us extend an invitation to numerical physics to reopen the study of BP and related models, taking into account the present locality approach. We hope that this will allow matching the recent theoretical understanding of sharp thresholds in two dimensions, as well as BP and related models in higher dimensions, within the universality framework and beyond it. To mention a few possible directions, it would be interesting to investigate to what extent other models are close to their local counterparts, to determine the leading asymptotics of so-called unbalanced critical models (see \cite{Bollobas23}) or practically determine the rather implicit leading asymptotics of balanced models (see \cite{Duminil-Copin24}), as well as to explore even the standard models studied here in three spatial dimensions (see \cite{Balogh12}).

\begin{acknowledgments}
\paragraph{Acknowledgments}
We thank Persi Diaconis, Alexander Holroyd and Cristina Toninelli for helpful remarks and Paul Th\'evenin for enlightening discussions regarding \cite{ElveyPrice21}. I.H.~was supported by the Austrian Science Fund (FWF): \mbox{P35428-N}. A.T.~was supported by grants ``Projeto Universal'' (406250/2016-2) and ``Produtividade em Pesquisa'' (304437/2018-2) from CNPq and ``Jovem Cientista do Nosso Estado'', (202.716/2018) from FAPERJ.
\end{acknowledgments}

\bibliography{Bib}

\newpage
\onecolumngrid
\setcounter{page}{1}
\thispagestyle{empty}
\begin{center}
\textbf{\large Supplemental Material: Locality approach to the bootstrap percolation paradox}
\end{center}
This supplemental material documents the data generation and analysis discussed in the letter.\\

\twocolumngrid
\section{Exact computation for local densities}
In order to study the local infection model, one can make use of dynamic programming techniques to obtain the exact densities of critical droplets.
The precise transition rates used in the algorithm can be found in \cite[Table 1]{Hartarsky24locality} for FBP, while for MBP they can be read off the source code file \url{bootstrap/taichi/modified.py}, available at the repository \url{https://github.com/augustoteixeira/bootstrap}. The resulting data is gathered in Tab.~\ref{tab:data}. Besides the efficient algorithm we have employed, two technical ingredients were necessary for us to reach such low values of $p$.
First, we made use of a GPU to take advantage of the high parallelism present in the calculations.
We ran this code on a Intel Xeon E3-1240 v3, with a single NVIDIA GeForce RTX 4060 8GB during 24 hours.
Secondly, in order to deal with the very low numbers that appear in the calculations without losing precision, we store the logarithms of each probability ($a = \log(r)$), using the expression $\log(r + s) = \log(e^a + e^b) = \max\{a, b\} + \log \big(1 + e^{-|a - b|} \big)$, in order to obtain the log of the sum or two small probabilities without losing precision. Based on this data, we progressively build the expressions in \eqref{eq:FBP} and \eqref{eq:MBP} and subsequent terms, assuming no prior knowledge for a start, see the scaling discussions below.
The code for the various fits below can be found at \url{https://github.com/augustoteixeira/bootstrap} in \url{fitting.py}.

\begin{table}[b]
    \centering
    \begin{tabular}{c|c|c}
    {$\log_2 p^{-1}$}&{$\log\rho_\ell^{-1}(p)$, FBP}&{$\log\rho_\ell^{-1}(p)$, MBP}\\\hline
    {2}&{3.6462939089044335}&3.025003004824336\\
    {3}&{9.48534315586599}&6.778614767734161\\
    {4}&{24.785862065200995}&17.63216670792452\\
    {5}&{61.09464730696058}&45.66021724467772\\
    {6}&{142.44209408918184}&112.51140378895116\\
    {7}&{318.20988111558466}&263.82432820233373\\
    {8}&{689.051877876013}&594.4366647140112\\
    {9}&{1458.978748960122}&1299.3999937139088\\
    {10}&{3039.8354477597804}&2776.8561233741584\\
    {11}&{6260.721269989636}&5835.068010227808\\
    {12}&{12787.496049133362}&12108.37054514238\\
    {13}&{25962.723826269754}&24891.836236292278\\
    {14}&{52486.88654620641}&50814.67137292014\\
    {15}&{105782.68428205681}&103192.960947481\\
    {16}&{212726.28468173486}&208743.86987754496\\
    {17}&{427113.5701763713}&421026.7811793454
    \end{tabular}
    \caption{Numerical data for local critical droplet densities.}
    \label{tab:data}
\end{table}

\section{Monte Carlo simulation of non-local densities}
In order to contrast the results for the local model with those of the original Bootstrap Percolation dynamics, we have also performed Monte Carlo simulations of both FBP and MBP for moderately small values of $p$.
In Figure~\ref{fig:plots}, these results have been represented with blue circles.
The code for such simulations can be found in the folder \url{bootstrap/rust_mc/} of the repository \url{https://github.com/augustoteixeira/bootstrap} and it does not involve any refined optimisation, besides using a low level programming language and making use of various cores in parallel.
The results can be found in Tab.~\ref{tab:MC}.

\begin{table}[b]
    \centering
    \begin{tabular}{c|c|c}
    {$\log_{\sqrt[5]2}p^{-1}$}&{$\log\rho^{-1}(p)$, FBP}&{$\log\rho^{-1}(p)$, MBP}\\\hline
    {10}&1.66297868&0.88591602\\
    {11}&1.90880149&0.93774197\\
    {12}&2.31055326&1.20306221\\
    {13}&2.9839127&1.38709304\\
    {14}&3.7385026&1.50706354\\
    {15}&5.12426155&1.96544661\\
    {16}&6.8547289&2.71492423\\
    {17}&8.76065061&3.726459\\
    {18}&11.40863243&5.16653646\\
    {19}&14.59271272&7.1772021\\
    {20}&18.43952811&9.56943828\\
    {21}&&12.94289428
    \end{tabular}
    \caption{Monte Carlo data for non-local critical droplet densities.}
    \label{tab:MC}
\end{table}

\section{First order scaling}
We start by determining the power $\alpha$ such that $\log\rho_\ell\sim p^{-\alpha}$. While Fig.~\ref{fig:FBP} clearly suggests that $\alpha\approx 1$, we would like to be more quantitative. As a first approximation, the last three discrete derivatives of $\log\log\rho_\ell^{-1}$ against $\log p^{-1}$ are $1.011$, $1.008$ and $1.006$. The fact that they are close to 1, but varying suggests that an additional corrective term is needed. We attempt a four-parameter fit of the form $\log\log\rho_\ell^{-1}\sim \alpha\log p^{-1}+a+bp^{c}$, $c> 0$. We would like to use only large values of $1/p$, but need at least four of them. Using only the last five data points, for FBP, we obtain $\alpha\approx0.9999(0)$. Moreover, this value is stable if we decide to use a different set (or number) of data points instead. This indicates that we have found the correct functional form for the leading term. Moreover, even though our guess for a pure power second term is wrong for MBP (see \eqref{eq:MBP}), the above four-parameter fit still gives $\alpha\approx0.999(7)$, as compared to the last discrete derivative of $1.012$.

\section{First order constant}
We have determined that $\alpha=1$. We repeat the same fit with the remaining three parameters $\log\log\rho_\ell^{-1}\sim \alpha\log p^{-1}+a+bp^{c}$, $c> 0$ to find $a$. This leads to $a\approx3.290(3)$ and this is also stable with respect to the choice of data points. Based on this and the rigorous lower bound on $\rho_\ell$ from \cite{Aizenman88} it is natural to guess that $a=\pi^2/3\approx 3.28987$, which could have been done without the matching rigorous lower bound of \cite{Holroyd03}. The same approach gives $a\approx3.29(2)$ for MBP.

\section{Second order scaling}
Whether we guessed that the leading asymptotic term is $\pi^2/(3p)$ as above or we knew it from \cite{Holroyd03}, our next task is to identify the functional form of the second order term. We set $\tilde\rho=\log\rho^{-1}_\ell-\pi^2/(3p)$ for convenience and we restart our analysis with $\tilde\rho$ instead of $\log\rho_\ell^{-1}$. We start by examining the discrete derivatives of $\log\tilde\rho$. For FBP, the last three are $0.514$, $0.511$, $0.509$, while the middle one is $0.536$. In comparison, the corresponding values for MBP are $0.579$, $0.574$ and $0.570$, and the middle one is $0.609$. This suggests a corrective term once again. However, in view of \cite{Hartarsky23mod-2n}, it is important to allow a more general form for it. We perform a five parameter fit of the form $\log\tilde\rho\sim a\log p^{-1}+b\log\log p^{-1}+c+dp^{e}$, $e>0$. With so many parameters, we face a problem: there are many local minima corresponding to spurious fits and local optimisers either get trapped in one of them or fail to converge, depending on initialisation. We solve this issue by first using more data points when fitting, say the last 10, to get a rough estimate of the relevant fit. We then narrow down the possible bounds for the five parameters and use only the last six data points to obtain an accurate prediction of their values. Finally, we verify that the resulting parameter values do produce a very good fit (mean squared errors are of order $10^{-13}$). 

For FBP, this yields $a\approx0.497$ and $b\approx0.06$, which supports the correct conclusion that $a=1/2$ and $b=0$. Instead, for MBP, we obtain $a\approx0.496$ and $b\approx1.05$, confirming that $a=1/2$ and $b=1$. It should be noted that in this case optimisation is particularly hard, due to the presence of a slightly worse fit around $a\approx 0.508$, $b\approx0.71$. Its mean squared error is only about 3 times larger, still of order $10^{-13}$ but it fluctuates more with the choice of data points, leading us to prefer the correct parameter choice. As we will see, this difficulty is due to our incorrect choice of third order correction.

We further remark that, if we had allowed a similar five-parameter fit for the first order term, the additional logarithmic correction is rejected by choosing a parameter close to 0 for it, as in  the case of the second order term of FBP.

\section{Second order constant}
To determine the correct constant $c$ above, we simply fix the values $a=1/2$ and $b=0$ for FBP respectively and perform the fit on the remaining three variables. This leads to $\sqrt p\tilde\rho\sim 11.58(9)$ to be compared with the rigorous value $2\pi\sqrt{2+\sqrt2}\approx 11.6098$ from \eqref{eq:FBP}. 

For MBP, proceeding similarly for $b=1$ tends to prioritise values of $e$ close to 0. Indeed, as it is visible from Fig.~\ref{fig:MBP}, we rather expect a third term of the form $p^{-1/2+o(1)}$, which means $e=0$. This suggests that the four parameter fit $\sqrt p\tilde\rho\sim a\log p^{-1}+b+cp^{d}$, $d>0$ is more appropriate. This gives $a\approx 1.848(4)$, again in agreement with the conjectured value $\sqrt{2+\sqrt2}\approx 1.84776$ from \eqref{eq:MBP}.

\section{Third order scaling}
In order to investigate the third order term, it is convenient to define $\bar\rho=\tilde\rho-2\pi\sqrt{2+\sqrt 2}/\sqrt p$ for FBP and $\bar\rho=\sqrt{2+\sqrt2}\frac{\log p^{-1}}{\sqrt p}-\tilde\rho$ for MBP instead. For FBP, examining the discrete derivatives of $\log\bar\rho$ reveals that they go down from $0.224$ at $p=2^{-2}$, to $0.188$ at $p=2^{-11}$ and then up to $0.195$ at $p=2^{-16}$. Moreover, adding logarithmic corrections is unsuccessful, so we expect 
a pure power third order term with an exponent close to $1/5$. 

For MBP, as seen from Fig.~\ref{fig:MBP}, a constant with power law correction gives a good fit for $\bar\rho\sqrt p$. On the other hand, proceeding as above, we could verify that more exotic possibilities such as polylogarithmic or iterated logarithmic leading terms deteriorate the result. Consequently, we expect that $\bar\rho\sqrt p$ converges to a constant.

\section{Third order constant}
For FBP, already the straightforward third order constant obtained as $p^{1/5}\bar\rho$ evaluated at the last data point yields an exceptionally good fit throughout the entire range we have access to (see Fig.~\ref{fig:FBP}). We therefore focus on MBP and perform a three parameter fit $\bar\rho \sqrt p\sim a+bp^{c}$, $c>0$. Using the last four data points (again, the result is not sensitive to this choice). This gives $a\approx 6.539$, $b\approx-7.4(4)$ and $c\approx 0.31$. This also leads to a convincing fit for all available values of $p$ (see Fig.~\ref{fig:MBP}). Interestingly, the value of the exponent $c$ suggests a fourth order term in MBP with exponent very close to the one of the third order term of FBP.

\section{Non-locality}
We now explain in more detail the reason for comparing $\rho_\ell$ with $p^2\rho$ and assess non-locality based on the comparison between local and non-local simulations. Firstly, the samples displayed in Fig.~\ref{fig:droplet} are not specially selected, but simply the first ones found. The one in Fig.~\ref{fig:droplet:FBP} happens to be local in the sense that there exists an initially infected starting $1\times 1$ square for which the local dynamics yields the final $93\times93$ square. This suggests that many internally filled critical droplets are actually local. 

Let us assume for a moment that all internally filled critical droplets are local and use this assumption to argue that $p^2\rho\sim\rho_\ell$. In this case, the event that the droplet is internally filled is the union over starting points of the probability that the local dynamics started there produces the final square. Thinking of a typical starting point, this probability is more or less $\rho_\ell$, itself essentially equal to the probability of growing to infinity from this starting point. If the events corresponding to different starting points were disjoint and had probability exactly $\rho_\ell$, we would have $\rho=\Lambda^2\rho_\ell$. 

This is only approximately true for multiple reasons. Firstly, starting points close to the boundary are less likely to succeed, because their growth is constraint to go in a particular direction from early on. Secondly, events are not quite disjoint. Indeed, there are always at least three valid starting points contained in a $2\times 2$ square (see Fig.~\ref{fig:droplet:FBP}). Sometimes there may be more possible starting points, but rarely many and they are always clustered together (it is not hard to prove along the lines of \cite{Hartarsky24locality} that distant successful starting points are asymptotically unlikely). Recall that $\Lambda^2=4\log^2(1/p)/p^2$ is a bit bigger than $1/p^2$. We neglect the remaining polylogarithmic factor in order to balance out the two effects above, which are difficult to quantify, but certainly at most polylogarithmic as well. This leads to the desired $\rho_\ell\sim p^2\rho$, up to logarithmic corrections, which have negligible effect already at the largest scales accessible to Monte Carlo simulations.

The perfect fit already at moderate values of $p$ between $\rho_\ell$ and $p^2\rho$ visible in Fig.~\ref{fig:FBP} further confirms the correctness of this reasoning and suggests that in FBP, locality is even stronger than what is proved in the locality theorem. For comparison, we have applied the same procedure to MBP. Here, already Fig.~\ref{fig:droplet:MBP} displays two non-local moves necessary for internally filling the critical droplet: for creating a $6\times 9$ rectangle from a $4\times 7$ one and then a $10\times 13$ rectangle from a $8\times 11$ one. If, despite this warning of non-locality, we compare $p^2\rho$ and $\rho_\ell$ for MBP as in Fig.~\ref{fig:MBP}, we observe that the two are not getting closer as $p$ decreases, but on the contrary. Moreover, the plot suggests that $(\log\rho_\ell-\log\rho)\sqrt p$ does not decay to $0$. While the data seems insufficient to advance more precise predictions, let us point out that this is in agreement with the heuristics of \cite[Section~8.2]{Hartarsky24locality}. In any case, the evidence is clear that locality for MBP is much weaker and only arises once droplets reach some intermediate scale.
\end{document}